# Relation of Microstreams in the Polar Solar Wind to Switchbacks and Coronal X-ray Jets

Marcia Neugebauer[1] and Alphonse C. Sterling[2]
1. Lunar and Planetary Laboratory, University of Arizona, Tucson, AZ 85721, USA, mneugeb@cox.net
2. NASA/Marshall Space Flight Center, Huntsville, AL 35806, USA, alphonse.sterling@nasa.gov.

ABSTRACT

Ulysses data obtained at high solar latitudes during periods of minimum solar activity in 1994 and 2007 are examined to determine the relation between velocity structures called microstreams and folds in the magnetic field called switchbacks. A high correlation is found. The possibility of velocity peaks in microstreams originating from coronal X-ray jets is re-examined; we now suggest that microstreams are the consequence of the alternation of patches of switchbacks and quiet periods, where the switchbacks could be generated by minifilament/flux rope eruptions that cause coronal jets.

## 1. INTRODUCTION

The Ulysses spacecraft sampled the solar wind emanating from polar coronal holes during periods of minimum solar activity from 1993 to 1996 and 2006 to 2008. Although the high-speed (> 700 km s$^{-1}$) flow was relatively steady, it showed some structures of solar origin such as occasional coronal mass ejections (Gosling et al. 1998), large amplitude Alfvén waves, and pressure-balance structures (Thieme et al. 1990; McComas et al. 1996). To that list, Neugebauer et al. (1995) added structures called *microstreams* in which the proton velocity was >20 km s$^{-1}$ above or below the running-average speed for an average duration of 0.4 days. Observations that (1) the microstreams were of temporal rather than spatial origin, (2) the velocity peaks had higher temperatures, greater ion fluxes, and greater relative numbers of ions with low first-ionization potentials than average, and (3) the velocity peaks were often bounded by discontinuities with large changes in the direction of the magnetic field led Neugebauer (2012) to suggest that polar X-ray jets were the source of the velocity peaks associated with the microstreams.

Balogh et al. (1999) identified intervals in the polar solar wind when the magnetic field underwent large directional changes with the radial component often reversing sign. Because the wave propagation direction reversed together with the field direction, these structures were identified as large-scale folds in the interplanetary field. Changes in the plasma flows during the magnetic reversals (Kahler at al., 1996; Yamauchi et al., 2004; Suess 2007; Neugebauer and Goldstein, 2013; Matteini et al., 2014) confirmed the nature of the direction changes as being folds rather than opposite polarity magnetic fields originating at the Sun. Such features are now called *switchbacks* (e.g., Yamauchi et al. 2004). With new observations available from the Parker Solar Probe (PSP), there has been renewed attention to switchbacks, including the finding that they occur in "*patches*" separated in time by quieter periods without switchbacks (Bale et al. 2019; Kasper et al. 2019; Horbury et al. 2020; Woodman et al. 2021). In a study of Alfvénic fluctuations in the polar solar wind Matteini et al. (2014) noted an association of switchbacks with local peaks in the proton speed. This paper provides a deeper study of that association.

Since 2012 when Neugebauer suggested that X-ray jets in the solar corona might be the source of microstreams, there have been additional studies of the origin and properties of X-ray jets. Moreover, Sterling and Moore (2020) have argued that such jets could be a source of solar-wind switchbacks.

Sections 2 and 3 of this paper examine the relation of microstream peaks to switchbacks. Then, in Section 4 we present the conclusion that microstream velocity peaks are caused by patches of switchbacks and we reconsider the possible connection between microstreams, switchbacks, and coronal jets.

2. SAMPLE ULYSSES OBSERVATIONS OF MICROSTREAMS AND SWITCHBACKS

Two of the microstreams observed by Ulysses are examined further to determine their relation to switchbacks. Here we present evidence that there is a strong connection between the two phenomena.

Figure 1 shows an example from 2007 December 3, where the top panel shows the proton velocity $V_p$ in km s$^{-1}$; the center panel shows the ratio of the radial component of the magnetic field $B_R$ to the magnitude B averaged over the 4 or 8 minutes required to obtain an ion spectrum; and the bottom panel shows $V_{ap}$, which is the difference between the alpha-particle speed $V_a$ and the proton speed $V_p$ in km s$^{-1}$, both calculated from the moments of the distribution functions. The vertical lines depict the boundaries of a microstream peak. In this case, the duration of the enhanced velocity was ~10.3 hours, and thus comparable to the typical durations of 0.4 days reported by Neugebauer et al. (1995). This event was detected when Ulysses was at solar distance of 1.8 AU and a heliographic latitude of 70.9°.

During the period between the vertical lines there are multiple peaks in $V_p$ that temporally correlate with reversals of the signs of $B_R/B$ and $V_{ap}$. In the normal fast solar wind from polar coronal holes, alpha particles are faster than the protons with $V_{ap} > 0$. Thus periods of a reversal of the sign of $V_{ap}$ would be consistent with a switchback (Yamauchi et al. 2004) if the magnetic field was folded at the same time. Ideally the magnitude of $V_{ap}$ would remain unchanged, but the data in Figure 1 are too noisy to test whether or not that is the case. The signs of $B_R/B$ and $V_{ap}$ are reversed for about one third of the time between the lines. Thus it appears that this microstream peak consisted of a series of switchbacks with durations on the order of roughly one hour. There is an indication of some smaller switchbacks before ~05:00, but there were none between 15:00 and 19:00. This suggests that the microstream peak might have been a consequence of an accumulation, or a "patch", of a series of switchbacks.

Figure 2 shows a second example of a microstream in the same layout as Figure 1. This microstream was detected on 2007 December 19, when Ulysses was at 1.91 AU and a heliospheric latitude of 76.2° and the solar field was again directed inward over the northern pole. The velocity enhancement lasted only ~3 hours, which is somewhat shorter than the average reported by Neugebauer et al. (1995). The correlated variations of $V_p$, $B_R/B$, and $V_{ap}$ indicate that this microstream peak was bounded by two switchbacks, one at its start and the other at its end. As with the microstream peak shown in Figure 1, there is some indication of switchback activity ahead of the peak, but shows no such activity in the following hours.

3. ADDITIONAL ULYSSES OBSERVATIONS

Further information about the relation between switchbacks and microstreams was obtained by examining data acquired by Ulysses during its 1994 passage through the flow from the southern solar coronal hole. Figure 3 is a plot of the proton speed between July 12 (-72.4° heliographic latitude and a solar distance of 2.73 AU) and October 28 (-72.8° and 1.98 AU). The extreme heliographic latitude of -80.2° was passed on September 13. The coronal field was pointed inward during this period with $B_R$ usually < 0. The red points in Figure 3, which constitute 12% of the total, denote proton velocities obtained when $B_R > 0$.

Almost all of the microstream peaks are capped by red dots and all of the microstream minima have black dots. Because the ambient field was inward during this period, the red dots are indicative of the presence of switchbacks. Thus the strongest peaks in proton velocity coincided with passage of switchbacks. This pattern is similar to the pattern observed closer to the Sun by PSP of periods with many switchbacks (i.e. patches) interspersed with periods of quiet field (Bale et al. 2019; Horbury et al. 2020; Woodham et al. 2021). In similar plots of shorter duration Matteini et al (2014) showed a correlation between switchbacks and microstream peaks during the Ulysses polar passages. Here we add consideration of the effect of alternating periods of switchback patches and quieter periods to yield the microstream peaks and dips, respectively

The Mozer et al. (2020) study of switchbacks observed by PSP showed that the rotation angle of the magnetic field increased with distance from the Sun, with the fraction rotating >90° changing from <5% at 35 solar radii ($R_s$) to ~15% at 50 $R_s$. For the interval shown in Figure 3 the Parker spiral angle varied between 5.8° and 8.5°, so the red dots suggest rotations of ~80° or more, in agreement with the radial evolution found by Mozer et al.

Over the course of the period shown in Figure 3 the average Alfvén speed increased from ~35 to ~42 km s$^{-1}$ as the spacecraft drew closer to the Sun. One would expect the velocity within a switchback to exceed the background velocity by approximately the Alfvén speed and thus also increase the spread between the red and black dots in Figure 3, which does appear to be the case.

## 4. DISCUSSION, AND A POSSIBLE X-RAY JET CONNECTION

We have shown that microstream peaks frequently coincide with structures that are identifiable as switchbacks based on $V_p$, $B_R/B$, and $V_{ap}$ measurements. The data are consistent with velocity peaks being a consequence of several switchbacks occurring in relatively close succession in time. Observations of switchbacks from PSP show that they often occur in intermittent patches lasting hours to days, and this could be consistent with switchbacks grouping together to form microstreams.

Neugebauer (2012) suggested that solar polar X-ray jets were the source of the velocity peaks associated with the microstreams. Here, we briefly reconsider the possible connection between coronal jets and microstreams in light of recent jet studies.

Coronal jets are long (~50,000 km), narrow (~10,000 km), transient (~10s of min) features observed in soft X-ray and EUV images (for recent reviews, see Raouafi et al. 2016; Hinode Review Team et al. 2019; Shen 2021). Recent ideas suggest that they result from eruptions of "minifilaments," analogous to the larger-scale filament eruptions that make typical solar flares and CMEs (Sterling et al. 2015). Frequently, the minifilaments erupt at sites of magnetic flux cancellation (e.g., Panesar et al. 2016a). Sterling & Moore (2020) suggest that miniature, twisted flux ropes that form in this manner and erupt as minifilaments to produce jets might transfer their twist onto open magnetic fields and propagate outward as Alfvénic disturbances that steepen into pulses that are detected as switchbacks in the solar wind.

So, could the coronal jets be the source of the microstreams, as suggested by Neugebauer (2012)? A potential difficulty is that it is not immediately clear how the differing time scales of jets (tens of minutes) could be reconciled with those of microstreams (several hours). Since most jets are observed to occur on time scales of <~1 hr, it is possible that cancellation continuing on several-hour time scales without erupting tends to cancel enough flux to make (mini)filament flux ropes large enough to erupt and make small CMEs rather than coronal jets (Sterling et al. 2018). Thus, hours-long-duration coronal jets, if they exist at all, may be too infrequent for a single such jet to account for a multi-hour-long microstream peak.

But if, as our studies in this paper indicate, microstreams might be the result of accumulated and persistent velocity enhancements resulting from a series of switchbacks, then it could be that individual switchbacks result from coronal jets, and the microstreams are a consequence of a series of such jet-driven switchbacks occurring in close succession. Thus, this would be a modification of the idea put forth by Neugebauer (2012) whereby a series of minifilament eruptions capable of producing coronal jets could accumulate and generate a microstream. In fact, homologous jets, continuing for hours at a time, have been commonly observed (e.g, Chifor et al. 2008; Cheung et al. 2015; Panesar et al. 2016a, 2016b; Sterling et al. 2017; Joshi et al. 2017; Paraschiv & Donea 2019, Moore et al. 2021). Under the minifilament-eruption scenario, the multiple minifilament/flux-ropes would be ejected as long as the cancelation continues (Panesar et al. 2016a, Sterling et al. 2017). A swarm of such homologous jets, produced over a several-hour time period, conceivably could account for a microstream peak. Additionally, there is some recent evidence (Bale et al. 2021; Fargette et al. 2021) that switchbacks have an extent similar to the scale size of supergranules (~30,000 km). Measurements of the lengths of the erupting minifilaments that produce jets range from ~8000 km (Sterling et al. 2015) to ~18,000 (Panesar et al. 2016), and thus of similar order to (albeit somewhat smaller than) a typical supergranule diameter. Fargette et al. (2021) also found switchbacks to occur on another sized scale, one that approximately corresponds to the size of photospheric granules, ~1000 km. Chromospheric spicules have widths of some fraction of this size, and thus their observation could be consistent with some spicules resulting from the minifilament-eruption-jet mechanism as suggested in Sterling & Moore (2016), and then those spicules making smaller-scale switchbacks as suggested in Sterling & Moore (2020).

This view does not rule out that sources other than coronal jets resulting from erupting minifilament/flux ropes might be responsible for the switchbacks. It would be useful to trace back microstreams observed in the solar wind to their source locations on the Sun, and to see whether they correspond to episodes of homologous and/or close-proximity nearly concurrent jets. Such studies, however, would best be performed with data acquired much closer to the Sun than the orbit of Ulysses allowed. One approach to testing the ideas presented here would be to try to trace back the source of one or more larger PSP-observed switchback patches to the source on the Sun. In this regard, a useful tool would be a solar wind model that extends from the solar surface out to the heliosphere (e.g., Lionello et al. 2016; Roberts et al. 2018), in which a solar wind disturbance can be driven and followed by modeling the magnetic field of an erupting minifilament.

We thank R. L. Moore and S. T. Suess for helpful comments and discussions. All the data used in this study are available from NASA's Space Physics Data Facility (cdaweb.gsfc.nasa.gov). ACS was supported by funding from the Heliophysics Division of NASA's Science Mission Directorate through the Heliophysics Guest Investigators (HGI, grant number 17-HGISUN17_2-0047) Program, the Heliophysics System Observatory Connect (HSOC, grant number 80NSSC20K1285) Program, and the MSFC Hinode Project.

ORCID iDs
M. Neugebauer: https://orcid.org/0000-0003-3766-4911
A. C. Sterling: https:/orcid.org/0000-0003-1281-897X

Figure captions

Figure 1. Variations of the proton speed $V_p$, the radial component of the magnetic field $B_R$ relative to the field magnitude B, and $V_{ap} = V_a - V_p$, which is the differential flow between protons and alpha particles across the microstream peak observed on 2007 December 3.

Figure 2. Variations of the proton speed $V_p$, the radial component of the magnetic field $B_R$ relative to the field magnitude B, and $V_{ap} = V_a - V_p$, which is the differential flow between protons and alpha particles across the microstream peak observed on 2007 December 19.

Figure 3. Proton speed observed by Ulysses between July 12 and October 28, 1994. Red and black dots denote times for which the radial component of the interplanetary magnetic field was outward or inward, respectively.

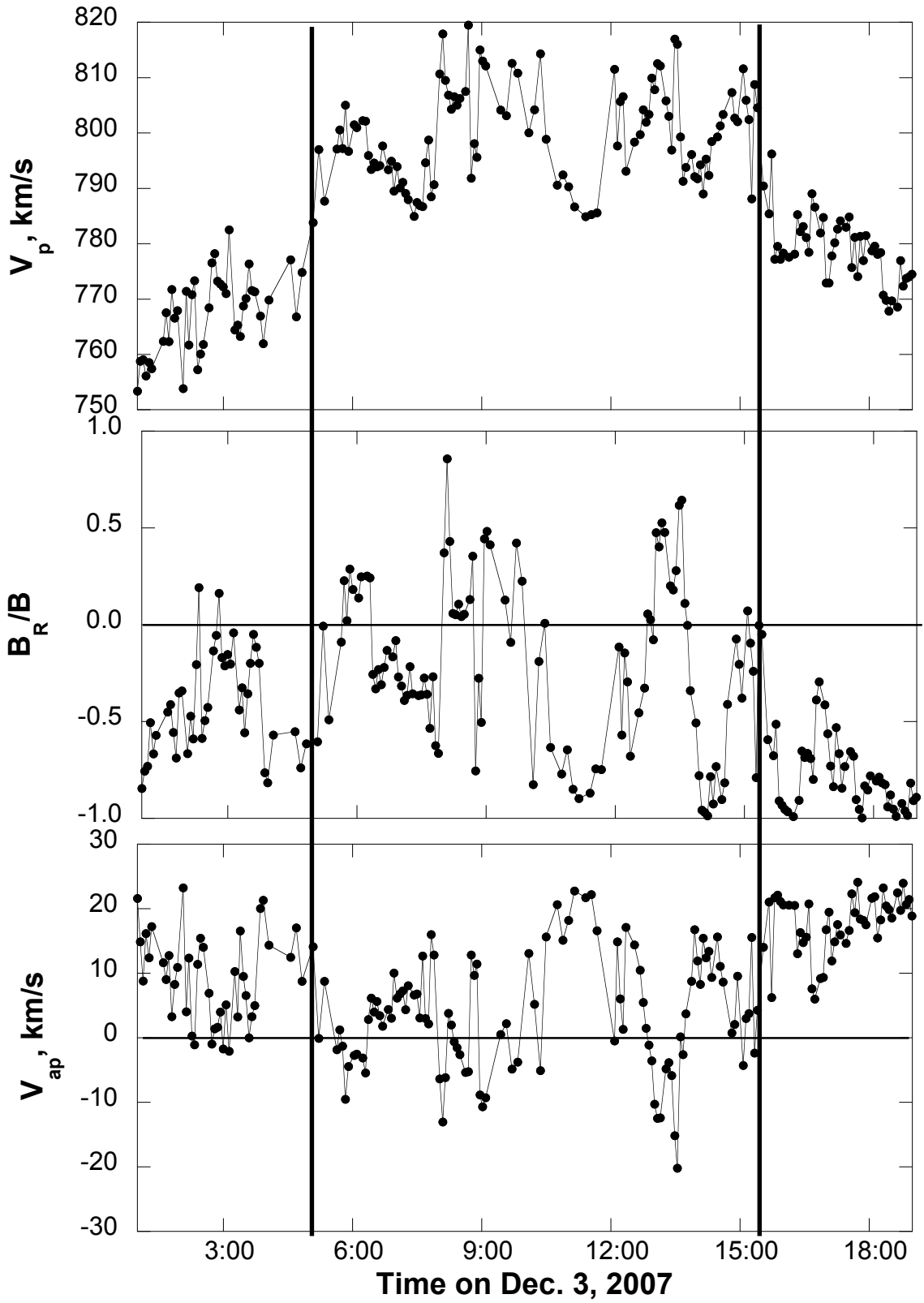

**Figure 1**

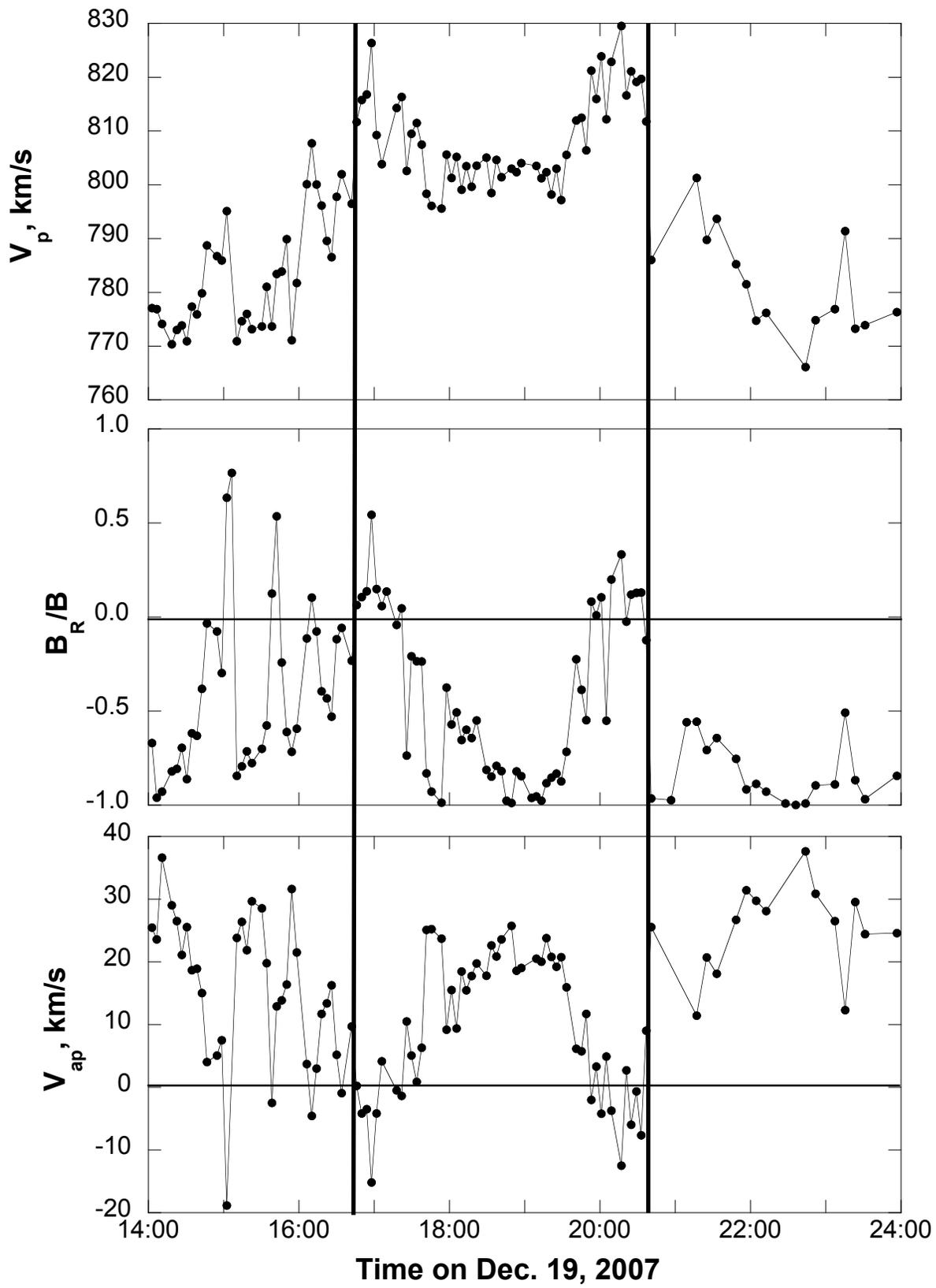

**Figure 2**

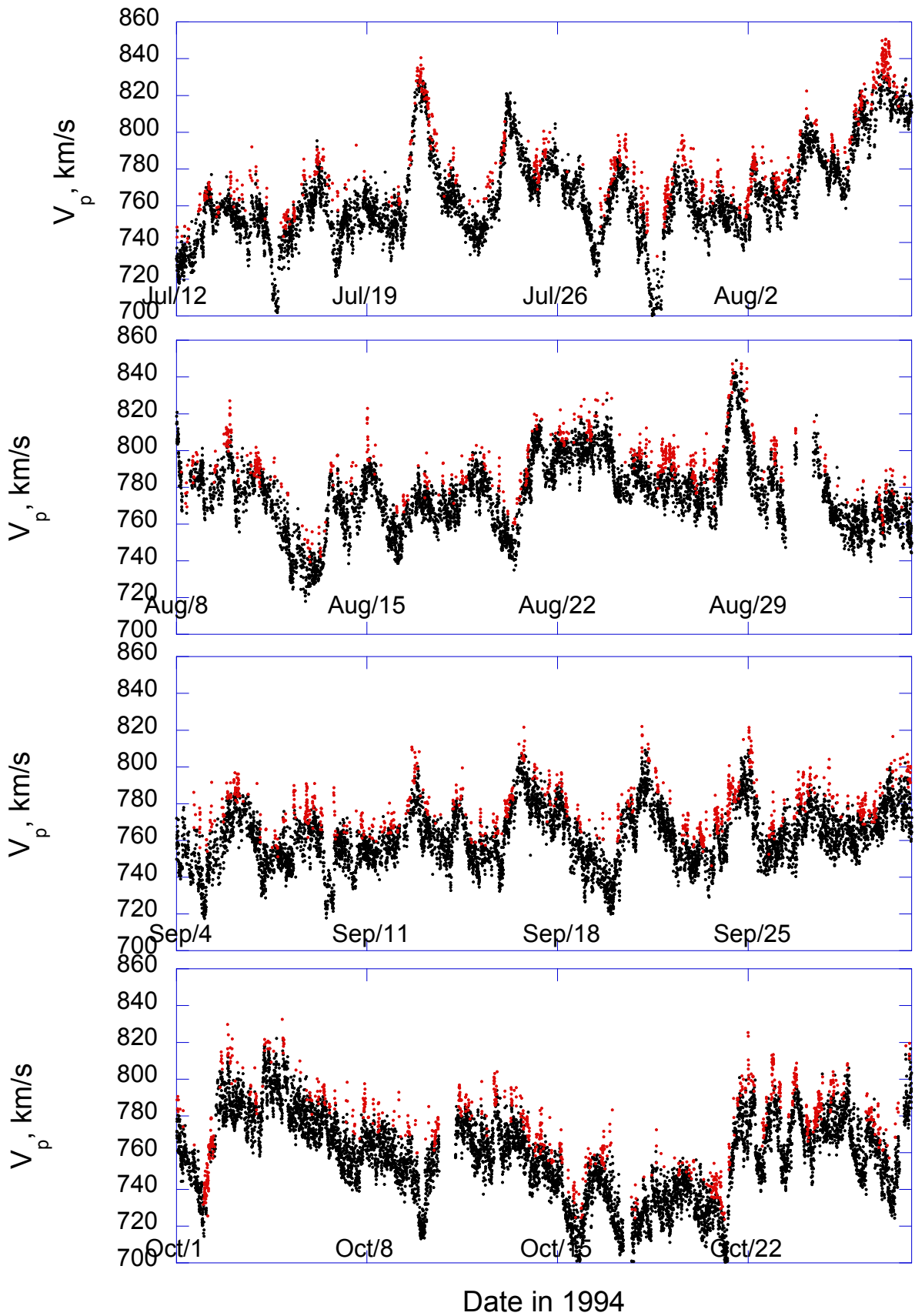

**Figure 3**